\documentclass[vecphys]{svmult}

\usepackage{makeidx}         
\usepackage{graphicx}        
\usepackage{multicol}        
\usepackage[bottom]{footmisc}

\makeindex             

\newcommand{\ha}{{H}{$\alpha$}}
\newcommand{\sini}{\ensuremath{\sin{i}}}

\newcommand{\pbp}{\ensuremath{\phi_b}}
\newcommand{\td}{\ensuremath{\theta}}
\newcommand{\tb}{\ensuremath{\theta_b}}
\newcommand{\Vmod}{\ensuremath{V_{\rm model}}}
\newcommand{\Vsys}{\ensuremath{V_{\rm sys}}}
\newcommand{\Vrot}{\ensuremath{\bar V_t}}
\newcommand{\Vrad}{\ensuremath{\bar V_r}}
\newcommand{\Vbit}{\ensuremath{V_{2,t}}}
\newcommand{\Vbir}{\ensuremath{V_{2,r}}}

\newcommand{\Vrotr}{\ensuremath{\Vrot(r)}}

\newcommand{\Vbitr}{\ensuremath{\Vbit(r)}}
\newcommand{\Vbirr}{\ensuremath{\Vbir(r)}}

\newcommand{\rotrad}{radial}
\newcommand{\rotbi}{bisymmetric}

\begin{document}

\title*{Modeling Non-Circular Motions in Disk Galaxies: A Bar in NGC~2976}
\titlerunning{Non-Circular Motions in Disk Galaxies}
\author{K. Spekkens\inst{1,2}\and J. A. Sellwood\inst{2}}
\authorrunning{Spekkens \& Sellwood}
\institute{National Radio Astronomy Observatory (NRAO)\footnote{NRAO is a facility of the National Science Foundation operated under cooperative agreement by Associated Universities, Inc.}
\and Department of Physics and Astronomy, Rutgers, the State University of New Jersey, 136 Frelinghuysen Road, Piscataway, NJ, USA 08854 \texttt{spekkens@physics.rutgers.edu, sellwood@physics.rutgers.edu}}
%
%
\maketitle

\abstract
 We give a brief description of a new model for non-circular motions in disk galaxy velocity fields, that does not invoke epicycles. We assume non-circular motions to stem from a bar-like or oval distortion to the potential, as could arise from a triaxial halo or a bar in the mass distribution of the baryons. We apply our model to the high-quality CO and \ha\ kinematics of NGC~2976 presented by \cite{simon03}; it fits the data as well as their model with unrealistic radial flows, but yields a steeper rotation curve. Our analysis and other evidence suggests that NGC~2976 hosts a bar, implying a large baryonic contribution to the potential and thus limiting the allowed dark matter halo density.

\section{Introduction}
\label{intro}

 A robust determination of the mass distribution in a spiral galaxy from its kinematics requires an estimate of the contribution from non-circular motions to the observed flow pattern. With the availability of high-quality velocity fields for many galaxies in the local volume (e.g. Koribalski, Walter, this volume), detailed studies of the importance of non-circular motions in spirals are now feasible.

 The flow pattern of an observed velocity field is typically analysed by performing an harmonic analysis in each of a series of concentric rings (e.g. \cite{begeman87},~\cite{schoen97}). Standard practice is then to invoke perturbed epicycle theory to interpret the resulting kinematic components in terms of physical processes or structures in the disk (e.g. \cite{schoen97}). However, some nearby systems exhibit non-circular flows whose amplitudes rival that of the inferred mean orbital speed \Vrotr\ (e.g. \cite{simon03}): the epicyclic approximation breaks down in this case, and the mass distribution derived by assuming that \Vrotr\ reflects circular orbital balance in the system (ie. adopting \Vrotr\ as the ``rotation curve'') is suspect.

 We recently presented a new technique for fitting generalized non-axisymmetric flow patterns to velocity fields without making the epicyclic approximation  (\cite{ss07}). We summarize this work here, and focus on a model in which flows stem from a bisymmetric distortion to an axisymmetric potential (Sect.~\ref{model}). The model describes the kinematics of NGC~2976 presented by~\cite{simon03} as well as that from their harmonic analysis. We find a steeper rotation curve in NGC~2976 than estimated by these authors, which we attribute to a large baryonic contribution from a bar (Sect.~\ref{n2976}).

\begin{figure}[t]
\centering
\includegraphics[height=6cm,viewport=0 74 470 440,clip]{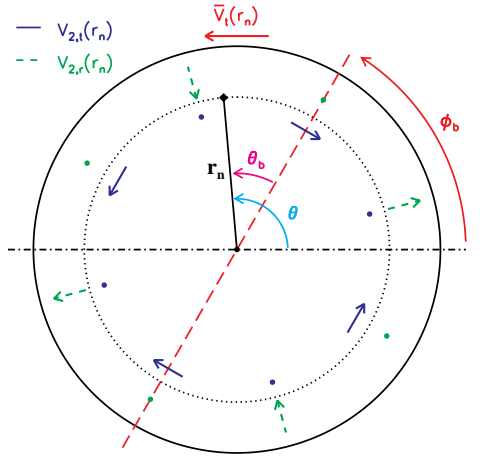}
\caption{Bisymmetric model flow pattern in the disk plane. The horizontal dash-dotted line is the major axis. The diamond denotes a point at $(r_n,\td)$, and the arrows show the extrema of \Vbir$(r_n)$ and \Vbit$(r_n)$. Adapted from Fig.~1 of \cite{ss07}. }
\label{setup}      
\end{figure}

\section{The Bisymmetric Model}
\label{model}

 We attempt to fit the non-circular motions in observed velocity fields by making the following assumptions:
\begin{itemize}
\item The non-circular flow stems from a bar-like or oval distortion to the potential.
\item Since the dominant kinematic signature of a bisymmetric distortion to the potential is also bisymmetric (e.g. \cite{sw93}), harmonics of order $m>2$ are neglected.
\item The non-circular motions are oriented about a fixed axis \pbp, with the radial and tangential components of the flow exactly out of phase with each other as required in a steady bar-like flow (e.g. \cite{sw93}).
\item The disk is flat, and thus has the same inclination $i$ at all radii $r$.
\end{itemize}

\begin{figure}[t]
\centering
\includegraphics[height=5.5cm,viewport=0 10 470 265,clip]{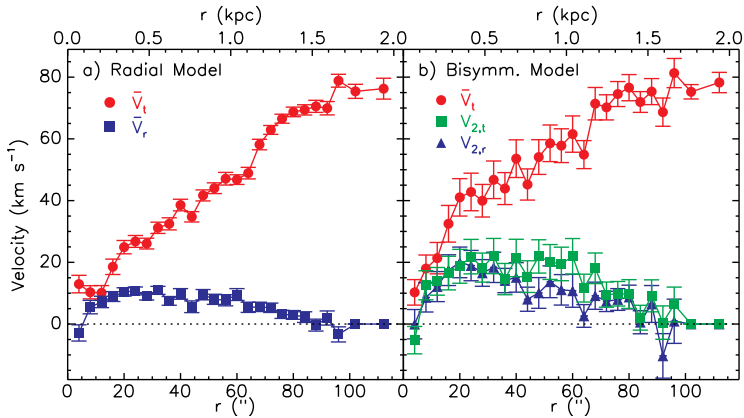}
\caption{Fitted velocity field components for NGC~2976 for the (a) \rotrad\ and (b) \rotbi\ models. Adapted from Fig.~3 of \cite{ss07}.}
\label{velcomps}      
\end{figure}

 The \rotbi\ model \Vmod$(r,\td)$ at some general point in the deprojected velocity field is therefore described by:
\begin{equation}
\frac{\Vmod - \Vsys}{\sini} =  \Vrot\cos{\td} - \Vbit\cos(2 \tb) \cos{\td}\, -
\,\Vbir\sin( 2 \tb) \sin{\td} \; ,
\label{bieq}
\end{equation}
and the geometry in the disk plane is shown in Fig.~\ref{setup}. The phases \td\ and \tb\ are the angles relative to the major axis and bisymmetric flow axis (or bar axis), respectively, and \Vsys\ is the disk systemic velocity. The non-parametric profiles \Vbirr\ and \Vbitr\ are the radial and tangential components of the bisymmetric flow. 

 For comparison with the work of \cite{simon03} we also consider a radial flow model, that is equivalent to a harmonic analysis including only $m=0$ terms:
\begin{equation}
\frac{\Vmod - \Vsys}{\sini} =  \Vrot \cos{\td} + \Vrad\sin{\td} \;.
\label{radeq}
\end{equation}

 When the non-circular motions are small, \Vrotr\ derived from (\ref{bieq})~and~(\ref{radeq}) will be comparable and will reflect circular orbital balance in the system: this is the essence of the epicyclic approximation (e.g. \cite{schoen97}). Since we do not require weak perturbations here, the different physical interpretations  invoked by the \rotbi\ and \rotrad\ models for the resulting flows may yield different \Vrotr\ as well. 

\section{A Bar in NGC~2976}
\label{n2976}
 
 We apply the models to the high-quality CO and \ha\ velocity field for the nearby, low-mass spiral NGC~2976 presented by \cite{simon03}. For each model we simultaneously fit for the velocity profiles and the disk properties, allowing each parameter to vary independently. 
 Our final \rotrad\ and \rotbi\ models of NGC~2976 have similar goodness-of-fit statistics, disk geometries and flow patterns: both are therefore adequate parametrizations of the data (see Table~1 and Fig.~2 of \cite{ss07}). 

 Figure~\ref{velcomps} shows the velocity field components obtained for both models. The radial model reproduces the results obtained by \cite{simon03}. While both models provide reasonable fits to the NGC~2976 kinematics, Fig.~\ref{velcomps} illustrates that \Vrotr\ from the \rotbi\ model rises more steeply than \Vrotr\ in  the \rotrad\ model.
  
\begin{figure}[t]
\centering
\includegraphics[height=4.7cm,viewport=0 75 470 255,clip]{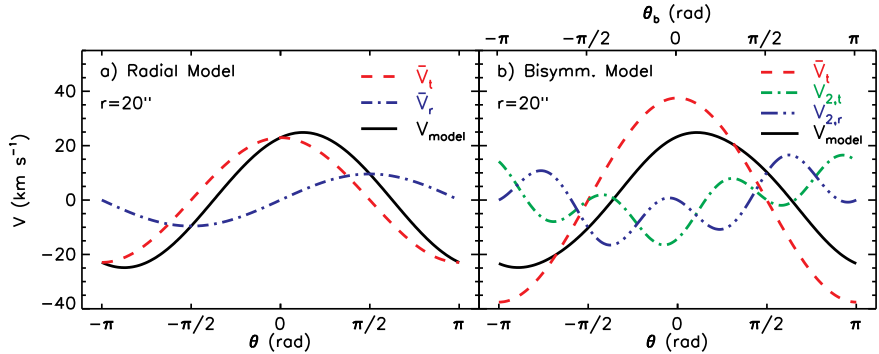}
\caption{Projected contributions from different kinematic components at $r=20''$  in the (a) \rotrad\ and (b) \rotbi\ models. Adapted from Fig.~6 of \cite{ss07}.}
\label{discuss}      
\end{figure}

 The reason for this difference is illustrated in Fig.~\ref{discuss}, that  shows sky-plane projections of the angular variations of the separate velocity profiles at $r=20''$ in both models. By definition, the non-circular motions in the \rotrad\ model must project to zero along the major axis [$\td\ = 0$]. This is not the case in the \rotbi\ model, where the contribution from non-circular motions depends on the phase of the bar. The bisymmetric model for NGC~2976 favors $\pbp \sim 17^{\circ}$, and in projection its bar and major axes are nearly aligned (see top horizontal axis of Fig.~\ref{discuss}). There is therefore a large negative contribution from \Vbit\ that offsets \Vrot\ at $\td = 0$, resulting in a \Vmod\ similar to that in the radial model.
 
  The amplitude of the radial flows required by the \rotrad\ model of NGC~2976 is prohibitively large (see \cite{ss07}). We therefore conclude that the \rotbi\ model better reflects the physical structure of this system. When non-circular motions are large \Vrotr\ is not a precise indicator of circular orbital balance, and we are developing self-consistent fluid-dynamical models of NGC~2976 to better measure this quantity. Nonetheless, the \rotbi\ model implies that the rotation curve of NGC~2976 rises more steeply than estimated by \cite{simon03}. There is evidence that the non-circular flows in NGC~2976 are caused by a bar rather than a triaxial halo: our kinematic bar axis and extent are identical to the photometric ones proposed by \cite{menendez07}, and \pbp\ is roughly coincident with the major axis of the CO distribution from \cite{simon03}. Since this in turn implies a large baryonic contribution to the potential, it is unlikely that the steeper rotation curve in NGC~2976 implies a high dark matter halo density.
  
  Even if the large non-circular motions found in other nearby spirals can be attributed to bar-like distortions, Fig.~\ref{discuss} shows that the \rotbi\ model will not always yield a steeper rotation curve than estimated previously. The change in \Vrotr\ relative to the \rotrad\ model result depends on the phase of the bar axis, and would be shallower than the latter if, for example, \pbp\ is aligned with the disk minor axis.

\vspace{0.05in}
We thank Josh Simon for providing the data for NGC~2976. KS is a Jansky Fellow of NRAO, and JAS is supported by grants AST-0507323 and NNG05GC29G.




%
%
%
%
%
%
%
%

%
%

%
%

%
%



\printindex
\end{document}